\documentclass[aps,amssymb,prl,twocolumn,superscriptaddress,showpacs]{revtex4}
\usepackage{graphicx,color}

\begin{document}

\title{Magnetic heat conductivity in $\rm\bf CaCu_2O_3$: linear temperature dependence}

\author{C. Hess}
\email[]{c.hess@ifw-dresden.de} \affiliation{Leibniz-Institute for Solid State and Materials Research, IFW-Dresden,
01171 Dresden, Germany}
%\author{P. Ribeiro}
\author{H. ElHaes}

\affiliation{2. Physikalisches Institut, RWTH-Aachen, 52056 Aachen, Germany and\\
Physics Department, Faculty of Women, Ain Shams University, Cairo, Egypt}
\author{A. Waske}
\affiliation{Leibniz-Institute for Solid State and Materials Research, IFW-Dresden, 01171 Dresden,
Germany}
% \affiliation{Leibniz-Institute for Solid State and Materials Research, IFW-Dresden, 01171 Dresden, Germany}
\author{B. B\"uchner}
\affiliation{Leibniz-Institute for Solid State and Materials Research, IFW-Dresden, 01171 Dresden, Germany}
\author{C. Sekar}
\affiliation{Leibniz-Institute for Solid State and Materials Research, IFW-Dresden, 01171 Dresden, Germany}
\author{G. Krabbes}
\affiliation{Leibniz-Institute for Solid State and Materials Research, IFW-Dresden, 01171 Dresden, Germany}
\author{F. Heidrich-Meisner}
\affiliation{Materials  Science and Technology Division, Oak Ridge National Laboratory,
 Oak Ridge, Tennessee, 37831, USA and\\
 Department of Physics and Astronomy, University of Tennessee, Knoxville,
 Tennessee 37996, USA}
\author{W. Brenig}
\affiliation{Institut f\"{u}r Theoretische Physik, Technische
Universit\"{a}t Braunschweig, 38106 Braunschweig, Germany}

\date{\today}

\begin{abstract}
We present experimental results for the thermal conductivity $\kappa$ of the pseudo 2-leg ladder material $\rm CaCu_2O_3$. 
The strong buckling of the ladder rungs renders this material a good approximation to a $S=1/2$ Heisenberg-chain. 
Despite a strong suppression of the thermal conductivity of this material in all crystal directions due to inherent disorder, we find a dominant magnetic contribution $\kappa_\mathrm{mag}$ along the chain direction. 
 $\kappa_\mathrm{mag}$ is \textit{linear} in temperature, resembling the low-temperature limit of the thermal Drude weight $D_\mathrm{th}$ of the $S=1/2$ Heisenberg chain. The comparison of $\kappa_\mathrm{mag}$ and $D_\mathrm{th}$ yields a magnetic mean free path of $l_\mathrm{mag}\approx 22\pm5$~\AA, in good agreement with magnetic measurements.
\end{abstract}

%\keywords{}

\maketitle

Recently, the \textit{magnetic} heat transport of low-dimensional quantum spin systems has become the focus of numerous studies
\cite{Sologubenko00,Hess01,Kudo01,Hess04a,Sologubenko00a,Sologubenko01,Ribeiro05,Hess03,Hofmann03,Sologubenko03,Sologubenko03a,Zotos97,Alvarez02,Klumper02,Heidrich02,Shimshoni03,Zotos04,Saito03,Orignac03,Rozhkov05,Jung06,Hess06} 
since intriguing properties have been found. 
Firstly, a substantial magnetic contribution to the thermal conductivity $\kappa$ in addition to a regular phononic background has been experimentally established for spin chain and ladder compounds as well as
two-dimensional (2D) antiferromagnets such as the insulating parent compounds of superconducting cuprates. It can thus be considered a generic feature of quasi low-dimensional magnetic materials.
Secondly, the surprisingly large magnetic contribution to $\kappa$ of spin ladder materials
observed in the case of  $\rm (Sr,Ca,La)_{14}Cu_{24}O_{41}$ \cite{Sologubenko00,Hess01,Kudo01,Hess04a}
 has triggered 
extensive theoretical work on possible ballistic heat transport in spin chains and
ladders \cite{Alvarez02,Klumper02,Heidrich02,Zotos04,Saito03,Orignac03,Jung06}. 

The analysis of experimental data for $\kappa$ can be quite involved, in particular 
when phonon and magnetic energy scales do not separate as is the case for the spin chain compounds 
$\rm Sr_2CuO_{3}$ and $\rm SrCuO_{2}$ \cite{Sologubenko00a,Sologubenko01}.
The situation in the cases of  spin ladder materials and the two-dimensional cuprates is quite fortunate:
here, a clear separation of phonon ($\kappa_\mathrm{ph}$) and magnetic ($\kappa_{\mathrm{mag}}$) contributions allows for a robust determination of $\kappa_{\mathrm{mag}}$
itself. It is important to note that $\kappa_{\mathrm{mag}}$ of the spin ladder compounds $\rm (Sr,Ca,La)_{14}Cu_{24}O_{41}$ \cite{Sologubenko00,Hess01,Kudo01,Hess04a}
shows thermally activated behavior at low temperature $T$ dominated by the large spin gap of two-leg ladders \cite{Sologubenko00,Hess01}, while in the case of 
La$_2$CuO$_4$ \cite{Hess03} $\kappa_{\mathrm{mag}}$ is found to be proportional to $T^2$, which is the leading contribution in $T$ to the specific heat of
a 2D square lattice antiferromagnet. Thus in these two cases, the experimentally observed $\kappa_{\mathrm{mag}}$ clearly exhibits 
intrinsic properties of the underlying spin models.

Theoretically, a consistent picture for thermal transport in spin-$1/2$ Heisenberg chains has only recently emerged: the integrability of this model
results in a divergent $\kappa_{\mathrm{mag}}$ \cite{Zotos97,Klumper02,Heidrich02} which is described by the so-called thermal Drude weight
$D_{\mathrm{th}}$ multiplied by a delta function at zero frequency. At low $T$ this $D_{\mathrm{th}}$ depends linearly on $T$ \cite{Klumper02} and such behavior
is  more generally expected for $\kappa$ of any spin chain model with gapless excitations \cite{Kane96,Heidrich02}. 

It is  the purpose of this Letter to present the first experimental example of a quasi one-dimensional (1D) quantum magnet (namely $\rm CaCu_2O_3$) which exhibits a {\it linear} $T$-dependence of $\kappa_{\mathrm{mag}}$ over a wide $T$-range. 
This is an eye-catching result as it resembles the intrinsic transport properties of a spin chain.
The extraction of $\kappa_\mathrm{mag}$ 
from the experimental data is very accurate since $\kappa_\mathrm{ph}$ of this material is strongly suppressed due to inherent disorder. 
We utilize well-known expressions for $D_\mathrm{th}$ to extract information on the scattering processes. Our analysis yields a $T$-independent magnetic mean free path $l_\mathrm{mag}\approx22\pm5$~{\AA}. The mean separation of magnetic defects along the chain direction as determined from magnetization measurements is of the same order of magnitude.

%%%%%%%%%%%%%%%%%%Properties%%%%%%%%%%%%%%%%%%%

The space group of $\rm CaCu_2O_3$ is $Pmmn$ with lattice constants $a=9.946$~{\AA}, $b=4.079$~{\AA}, and $c=3.460$~{\AA} \cite{Kim03}. 
The structure basically consists of $\rm Cu_2O_3$ structural units arranged in the $ab$-plane with the geometrical form of buckled two-leg ladders 
running along the $b$-direction. The $\rm Cu_2O_3$-planes are stacked along the $c$-direction and are separated from each other by layers of Ca-ions. 
The nearest neighbor magnetic exchange coupling of the $\rm Cu^{2+}$ spins along the 180$^\circ$ Cu-O-Cu bonds which form the ladder legs in the $b$-direction 
is large and has been estimated as $J/k_B\approx2000$~K \cite{Kiryukhin01}. The intra-ladder magnetic rung-coupling along the buckled Cu-O-Cu bonds ($123^\circ$ bonding angle) 
in the $a$-direction is much smaller and believed to be in the range $J_\perp/k_B\approx100-300$~K \cite{Kiryukhin01,Kim03,Goiran06}. 
The inter-ladder coupling along the $c$-axis has been estimated to be of the same order of magnitude \cite{Kiryukhin01,Kim03}. 
Along the $a$-axis a weak ($<100$~K) and frustrated nearest neighbor inter-ladder coupling is expected \cite{Kiryukhin01} and an even weaker ($8-30$~K) 
second nearest neighbor inter-ladder coupling has been suggested to be mediated via excess $\rm Cu^{2+}$ ions on interstitial positions \cite{Goiran06}. 
Despite $\rm CaCu_2O_3$ being a ladder-like material, where in principle a gapped nonmagnetic ground state is expected \cite{Barnes93}, the material orders 
antiferromagnetically at $T_N\approx25$~K \cite{Kiryukhin01,Goiran06,Sengupta04} and inelastic neutron scattering reveals a spin chain-like excitation 
spectrum with the upper bound for a spin gap $\Delta\approx3$~meV \footnote{B.~Lake et al., to be published.}. We therefore consider $\rm CaCu_2O_3$ as an ensemble of weakly 
coupled spin chains.

%{\em Experimental.---}
We have grown single crystalline $\rm CaCu_2O_3$ by the traveling solvent floating zone method \cite{Sekar05}. Samples with typical dimensions of around 3.5~mm length 
along the thermal current direction and 1~mm$^2$ cross section were cut from a crystal and the thermal conductivity of $\rm CaCu_2O_3$ has been measured along the 
$a$, $b$ and $c$-direction ($\kappa_a$, $\kappa_b$ and $\kappa_c$) as a function of $T$ in the range 7-300~K. We used a standard four probe technique where errors 
due to radiation loss are minimized. Magnetization has been measured using a superconducting quantum interference device (Quantum Design MPMS XL5).

\begin{figure}
\includegraphics [width=\columnwidth,clip] {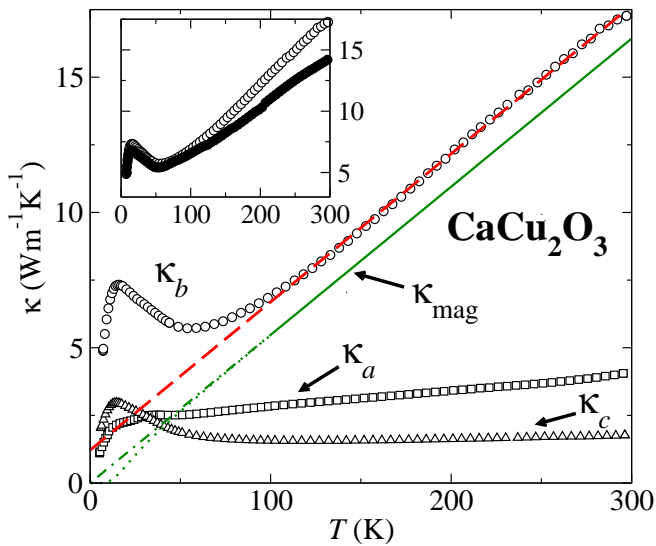}
\caption{\label{kappadat}$\kappa_a$ ($\square$), $\kappa_b$ ($\bigcirc$) and $\kappa_c$ ($\triangle$) of $\rm CaCu_2O_3$ as a function of $T$. The dashed and solid lines represent a linear fit of the experimental data in the range 100-300~K and the estimated $\kappa_\mathrm{mag}$ 
in this range. Extrapolations of $\kappa_\mathrm{mag}$ towards low $T$ (assuming a $T$-independent
$l_\mathrm{mag}$ as extracted for $T>100$~K) corresponding to a finite ($\Delta=3$meV) and a vanishing spin gap are represented by  dotted and dashed-dotted lines, respectively. Inset: $\kappa_b$ from two different measurements 
(open symbols depict the same curve as in the main panel).}
\end{figure}
 
% {\em Results.---}
Fig.~\ref{kappadat} shows $\kappa_a$, $\kappa_b$ and $\kappa_c$ of $\rm CaCu_2O_3$ as a function of $T$. Since the material is insulating, electronic 
heat conduction is negligible and we therefore expect these components to originate from phononic heat conduction plus a possible magnetic contribution. 
The thermal conductivities \textit{perpendicular} to the chain direction ($\kappa_a$ and $\kappa_c$) only exhibit a weak $T$-dependence and 
share absolute values ($\lesssim 4~\rm Wm^{-1}K^{-1}$) which are 1-2 orders of magnitude smaller than $\kappa_\mathrm{ph}$ 
of other chemically undoped cuprates such as $\rm SrCuO_2$ \cite{Sologubenko01} or $\rm Sr_{14}Cu_{24}O_{41}$ \cite{Hess01}. Instead of a pronounced phononic low-$T$ peak, which is usually found in such cases, only a small peak is present in $\kappa_c$ while no peak is found in 
$\kappa_a$. In fact, such strongly suppressed $\kappa$ is  typical for $\kappa_\mathrm{ph}$ with a  high phonon scattering rate \cite{Berman50}. 
Indeed, substantial phonon-defect scattering must be present in $\rm CaCu_2O_3$ due to inherent structural disorder induced by a significant Ca and oxygen deficiency being balanced by excess Cu \cite{Ruck01,Kim03}.
We therefore consider $\kappa$ perpendicular to the chains to be purely phononic. 

A completely different behavior is observed for $\kappa$ \textit{parallel} to the chains. At low $T\lesssim50$~K $\kappa_b$ resembles $\kappa_c$ although it is about three times larger. However, $\kappa_b$ steeply increases at $T\gtrsim50$~K with this increase being linear in $T$ for $T\gtrsim100$~K. Such a strong increase of $\kappa$ with rising $T$ cannot be understood in terms of conventional phonon heat conduction by acoustic phonons. Dispersive optical 
phonons could in principle give rise to an  increase of $\kappa_\mathrm{ph}$ with increasing $T$ \cite{Hess04} and possibly play a role in the $T$ dependence of $\kappa_a$. Heat transport by optical phonons is however unable to account for the rather large $\kappa_b$ at 300~K since this would require unrealistically large phonon mean free paths \footnote{A comparison with the structurally closely related doped $\rm La_2CuO_4$ \cite{Hess04} would suggest $l_\mathrm{ph}\approx270$~{\AA}.}. In analogy with observations in other cuprate materials \cite{Sologubenko00,Hess01,Kudo01,Hess04a,Sologubenko00a,Sologubenko01,Ribeiro05}, we therefore conclude that the strong increase of $\kappa_b$ and the resulting anisotropy of the $\kappa$ tensor originate from 1D heat transport due to magnetic excitations in the weakly coupled spin chains.
While these observations and the following analysis represent the central results of this paper, we wish to point out an additional feature of the thermal conductivity. As depicted in the inset of Fig.~1, we observe variations of $\kappa$ as measured during different runs which are beyond the statistical error of the typical measurement.
Similar has been observed in other 1D quantum magnets as well \cite{Sologubenko03,Sologubenkopriv,Uchidapriv,Hessunpub} and remains an open issue which merits further investigation.

% {\em Magnetic and phononic contributions.---}
In order to separate the phononic ($\kappa_{\mathrm{ph},b}$) and magnetic ($\kappa_\mathrm{mag}$) parts of $\kappa_b$ 
we assume that $\kappa_{\mathrm{ph},b}$ is of a similar magnitude and exhibits a similar $T$-dependence to the purely phononic $\kappa_a$ and $\kappa_c$. 
Possible errors due to the crudeness of this estimation become small at $T\gtrsim100$~K where the strongly increasing magnetic part of $\kappa_b$ becomes clearly larger than any possible phononic thermal conductivity. Since $\kappa_a$ and $\kappa_c$ are only weakly $T$-dependent and $\kappa_b$ increases linearly in this $T$-regime, it is natural to conjecture $\kappa_{\mathrm{ph},b}\approx\mathrm{const}$ and $\kappa_\mathrm{mag}\propto T$ at $T\gtrsim100$~K.
Indeed, a linear fit in the range 100-300~K (dashed line in Fig.~\ref{kappadat}) describes the data almost perfectly and yields $\kappa_{\mathrm{ph},b}=1.2\rm ~Wm^{-1}K^{-1}$ and $\kappa_\mathrm{mag}=0.055{\rm ~Wm^{-1}K^{-2}}\times T$. We plot the  extracted $\kappa_\mathrm{mag}$ for $T>100$~K as a solid line in Fig.~\ref{kappadat}. $\kappa_\mathrm{mag}$ at lower $T$ cannot be inferred from our data.
Returning to the inset we note that the alternative curve for $\kappa_b$ also 
increases linearly with $T$ for $T\gtrsim100$~K  yielding $\kappa_{\mathrm{ph},b}=2.2\rm ~Wm^{-1}K^{-1}$ and $\kappa_\mathrm{mag}=0.041{\rm ~Wm^{-1}K^{-2}}\times T$. 
We therefore assume  that the  difference in the two results for $\kappa_b$ represents the error in determining the intrinsic $\kappa$ of our sample.
$\kappa_{\mathrm{ph},b}$ could certainly exhibit a slight $T$-dependence, either increasing (as $\kappa_a$) or decreasing 
(as normally expected for $\kappa_{\mathrm{ph}}$) with rising $T$. These uncertainties are irrelevant to our further analysis.

% {\em Analysis of $\kappa_\mathrm{mag}$ and discussion.---}
Qualitatively, the linear increase of $\kappa_\mathrm{mag}$ with  $T$ is a remarkable result 
as it directly reflects the low-temperature behavior of the thermal Drude weight of a Heisenberg chain \cite{Klumper02,Heidrich02}, where (in SI units)
\begin{equation}
 D_\mathrm{th}=\frac{(\pi k_B)^2}{3\hbar}vT\,. \label{dmag}
\end{equation}
As the exchange integral $J/k_B$ along the chain is of the order of 2000 K, we may safely neglect any deviations
from the linear behavior that become relevant at temperatures $T\gtrsim 0.15 J/k_B$ \cite{Klumper02,Heidrich02}, i.e., $T\sim 300$ K.

As mentioned earlier, the structure of CaCu$_2$O$_3$ suggests a model of weakly coupled chains. We have checked that any 
deviations of $\kappa_{\mathrm{mag}}$ from a linear $T$-dependence due to the presence of a small spin gap can only occur
in the low temperature regime which is dominated by phonons and thus experimentally hardly observable. Using a Boltzmann-type expression for $\kappa_\mathrm{mag}$  (see Eq.(1) in Ref.~\cite{Hess01})  to estimate the effect of a small gap, 
we further find that at $T>100$ K a pure spin chain and a weakly coupled ladder for $J_\perp/J\lesssim0.05$ both result in the same linear $T$-dependence (cf. Fig.~\ref{kappadat}). For the latter, $\Delta\lesssim 3$~meV since $\Delta\approx 0.4J_\perp$ \cite{Johnston00}.
It is thus justified to use Eq.~(\ref{dmag}) to analyze our data at high $T$.

In the presence of external scattering, the thermal conductivity of a single chain $\tilde{\kappa}_\mathrm{mag}$ is rendered finite with a width $\sim1/\tau$ and 
 may be approximated by $\tilde{\kappa}_\mathrm{mag}= D_\mathrm{th} \,\tau /\pi$.
Combining this with Eq.~(\ref{dmag}) to calculate the magnetic mean free path $l_\mathrm{mag}=v\tau$ from our experimental data for $\kappa_\mathrm{mag}$ yields
\begin{equation}
l_\mathrm{mag}=\frac{3}{\pi}\frac{\hbar}{k_B^2N}\frac{\kappa_\mathrm{mag}}{T}\,,\label{lmag}
\end{equation}
where $N=4/ac$ is the number of spin chains per unit area in the crystal. 
A kinetic approach as described in Ref.~\onlinecite{Sologubenko01} 
yields the same result in the low temperature limit.
Note further that in our model, $\tau$ is an energy independent quantity which seems to be a reasonable approximation, since even at $T=300~\mathrm{K}\ll J/k_B$ only 
spinons in a small region of the Brillouin zone can contribute to $\kappa_\mathrm{mag}$. In this region of the Brillouin zone $v\approx Ja\pi/2\hbar$ is constant and hence $l_\mathrm{mag}$ can also be regarded as independent of energy.
From Eq.~\ref{lmag} and the experimental data we obtain $l_\mathrm{mag}=22\pm 5$~{\AA} for the entire range 100-300~K \footnote{The error in $l_\mathrm{mag}$ arises from the difference between the two curves in the inset of Fig.~\ref{kappadat}.}, corresponding to about $5-6$ lattice spacings.

Within the framework of Boltzmann-type kinetic models, $T$-independent mean free paths for magnetic excitations have already been observed in the spin ladder material 
$\rm (Sr,Ca,La)_{14}Cu_{24}O_{41}$ \cite{Sologubenko00,Hess01,Hess04a,Hess06} and the 2D antiferromagnet 
$\rm La_2CuO_4$ \cite{Hess03} where in the low-$T$ regime scattering off defects dominates over other possible scattering mechanisms. In analogy with these cases it seems reasonable to explain our result of a $T$-independent $l_\mathrm{mag}$ in $\rm CaCu_2O_3$ by 
dominant spinon-defect scattering as well. However, a constant $l_\mathrm{mag}$ over such a large $T$-range is very surprising because spinon-phonon scattering 
should become increasingly important at higher $T$ and eventually lead to a $T$-dependent mean free path. 
Despite this intuition the data 
suggest that at 300~K the probability of a spinon scattering off a phonon is still much lower than that of scattering off a defect. It is therefore natural to 
conclude that the density of relevant defects is very high in the material, resulting in scattering off defects being much more important than other 
scattering processes. This is consistent with our quantitative result for $l_\mathrm{mag}$ which is about 1-2 orders of magnitude smaller than in spin chain 
systems with similar magnetic exchange, e.g. $\rm SrCuO_2$ and $\rm Sr_2CuO_3$, where much larger and $T$-dependent mean free paths 
have been found \cite{Sologubenko00a,Sologubenko01}.

\begin{figure}
\includegraphics [width=\columnwidth,clip] {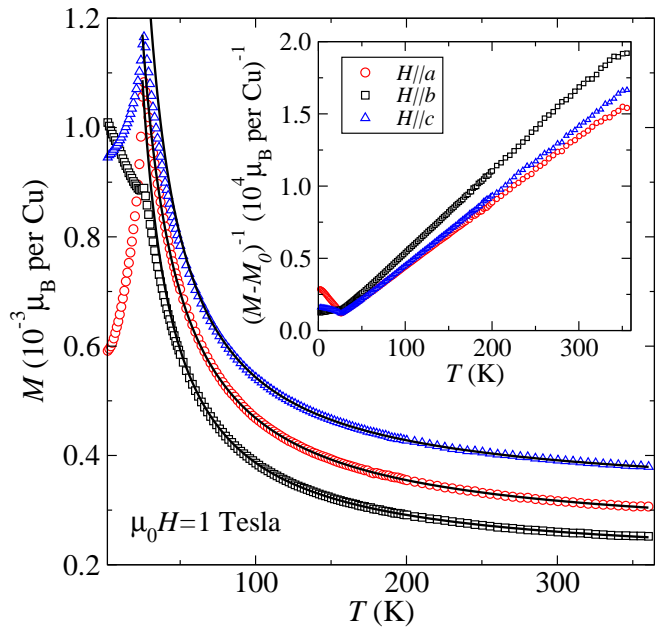}
\caption{\label{chidat}Magnetization of $\rm CaCu_2O_3$ as a function of temperature with a magnetic field $\mu_0H=1$~Tesla parallel to the three crystallographic axes. The solid lines represent Curie-Weiss-type fits to the data in the range 100-360~K which yield $\sim3$\% free spins with respect to Cu. Inset: Inverse magnetization after subtracting a constant $M_0$ which roughly accounts for van Vleck and chain magnetism \cite{Kiryukhin01}.}
\end{figure}

It is obvious to search for the origin of the high density of static scattering centers as suggested by our data in the strong off-stoichiometry of $\rm CaCu_2O_3$. The oxygen deficiency is clearly most relevant for creating vacancies within the $\rm Cu_2O_3$ chain structures which must create local scattering sites. In order to estimate the density of such scattering sites we consider 
the $T$-dependence of the magnetization $M$ of our sample, shown in Fig.~\ref{chidat}, which is in good agreement with previous results for crystals from a different sample growth \cite{Kiryukhin01,Goiran06}. $M(T)$ exhibits a sharp anomaly at the N\'eel temperature $T_N \approx 25$~K and a Curie-like behavior at $T>T_N$ indicating $\sim3$\% of free spins (with respect to Cu) in the material. 
Kiryukhin et al. suggested \cite{Kiryukhin01} that these free moments are located directly in the chains and originate from chain interruptions. According to their analysis the high-$T$ Curie tail in the magnetization originates from chain segments with a 
length between $\sim40$ and 80~{\AA}. In contrast, a recent electron spin resonance study by Goiran et al. \cite{Goiran06} suggests that the free moments are more likely to arise from excess Cu$^{2+}$ ions where each of these ions resides on an interstitial site in the vicinity of an oxygen vacancy in a neighboring chain structure. Within the latter scenario a lower limit for the density (per unit
length along the $b$-direction) of oxygen vacancies $n\gtrsim0.03/b$ can be inferred from the $M(T)$ data. This yields an upper limit for
the mean distance $d$ between the induced local scattering sites within a chain of $d=\frac{1}{2n}\lesssim 68$~{\AA}. In this estimate we take into account that each vacancy creates a local structural distortion which affects at least two chains since the individual chain structures are strongly interwoven.
In either case the extracted mean length of non-distorted chain segments $d$ should be regarded as an upper limit for the magnetic mean free path since not every chain distortion affecting $\kappa_\mathrm{mag}$ necessarily contributes to the magnetization. This agrees reasonably well with the analysis of the thermal conductivity as $l_\mathrm{mag}$ is found to be somewhat smaller (by a factor 2-3) than $d$.

We mention that the $T$-independent scattering rate as found in our experiment is in conflict with recent calculations of a scattering rate $\tau_\mathrm{imp}^{-1}\propto T^{-1}$ for impurities which induce slight disorder in the magnetic exchange coupling \cite{Rozhkov05}. A possible reason for this discrepancy could be related to the nature of the defects in our sample. Very likely the degree of the distortion of the chain at the actual scattering sites is too large for this model to be applicable. Detailed experimental and theoretical investigations are underway to clarify this issue.

% {\em Conclusion.---}
To conclude, we have studied the thermal conductivity of $\rm CaCu_2O_3$ as a function of temperature $T$.
The thermal conductivity parallel to the chains exhibits a pronounced linear increase with increasing $T$ which we attribute to magnetic heat transport within the chains of the material. The linear increase resembles the intrinsic thermal conductivity of a spin chain with a constant scattering rate. We extract a value for the magnetic mean free path $l_\mathrm{mag}\approx 22\pm5$~{\AA} which is in reasonably good agreement with the mean distance between magnetic defects in the material as determined from magnetization data.

%%%%%%%%%%%%%%%%%%%%%%%%%%%%%%%%%%%ACKNOWLEDGEMENTS%%%%%%%%%%%%%%%%%%%%%%%%%%%%%%%%%%%%%%%%
%{\em Acknowledgments.---}
It is a pleasure to thank A.V.~Sologubenko, X.~Zotos, A.L.~Chernyshev, S.-L.~Drechsler, V.~Kataev and R.~Klingeler for stimulating discussions and A.P.~Petrovic for proofreading the manuscript. F.H-M. is supported by NSF grant DMR-0443144.

%\bibliography{ladders}

\end{document}